\documentclass[aps,prd,preprintnumbers,superscriptaddress,nofootinbib,showpacs,twocolumn]{revtex4}%
\usepackage{amsmath,amssymb,bm,float,mathrsfs,subfigure,tabularx}
\usepackage{graphicx}
\usepackage{dcolumn}
\usepackage{color}
\usepackage{hyperref}
\usepackage{comment}

  \newcommand{\beq}{\begin{equation}}
  \newcommand{\eeq}{\end{equation}}
  \newcommand{\al}[1]{\begin{align} #1 \end{align}}
  \newcommand{\bi}{\begin{itemize}}
  \newcommand{\ei}{\end{itemize}}
  \newcommand{\bc}{\begin{center}}
  \newcommand{\ec}{\end{center}}
  
  \def\dd{\mathrm{d}}

  \def\rmm{\mathrm{m}}

  \def\mcM{\mathcal{M}}
  
  \def\mcO{\mathcal{O}}

  \def\pd{\partial}

  \newcommand{\ave}[1]{\left\langle #1 \right\rangle}

  \def\3Dint#1{\int\frac{\dd^{3}{#1 }}{(2\pi )^3}}

\begin{document}

\title{Constraining primordial non-Gaussianity via a multitracer technique
with surveys by Euclid and Square Kilometre Array}


\author{Daisuke Yamauchi}
\email[Email: ]{yamauchi"at"resceu.s.u-tokyo.ac.jp}
\affiliation{
Research Center for the Early Universe, Graduate School of Science, 
The University of Tokyo, Bunkyo-ku, Tokyo 113-0033, Japan
}

\author{Keitaro Takahashi}
\affiliation{
Faculty of Science, Kumamoto University, 2-39-1 Kurokami, Kumamoto 860-8555, Japan
}

\author{Masamune Oguri}
\affiliation{
Research Center for the Early Universe, Graduate School of Science, 
The University of Tokyo, Bunkyo-ku, Tokyo 113-0033, Japan
}
\affiliation{
Department of Physics, The University of Tokyo, Bunkyo-ku, Tokyo 113-0033, Japan}
\affiliation{Kavli Institute for the Physics and Mathematics of the Universe, 
The University of Tokyo, Kashiwa, Chiba, 277-8568, Japan}

\begin{abstract}
We forecast future constraints on local-type primordial non-Gaussianity
parameter $f_{\rm NL}$ with a photometric galaxy survey by Euclid, a
continuum galaxy survey by Square Kilometre Array (SKA), and their
combination. We derive a general expression for the covariance 
matrix of the power spectrum estimates of multiple tracers to show how
the so-called multitracer technique improves constraints on 
$f_{\rm NL}$. In particular we clarify the role of the overlap
fraction of multiple tracers and the division method of the
tracers. Our Fisher matrix analysis indicates that stringent constraints
of $\sigma (f_{\rm NL})\lesssim 1$ can be obtained even with a single
survey, assuming five mass bins. When Euclid and SKA phase 1 (2) are
combined, constraints on $f_{\rm NL}$ are improved to $\sigma (f_{\rm NL})= 0.61~(0.50)$.   
\end{abstract}

\preprint{}

\maketitle

\section{Introduction} 

Primordial non-Gaussianity of density fluctuations is key to
understanding the physics of the early Universe. 
Among several types of
primordial non-Gaussianity, the local-type one, $f_{\rm NL}$, has been
studied widely, partly because even the simplest inflationary models
predict small but nonvanishing values of $f_{\rm NL}$ of $\mcO (0.01)$.
Here we quantify non-Gaussianity of the local form as 
\al{
	\Phi =\phi +f_{\rm NL}(\phi -\ave{\phi^2})
	\,,
}
where $\Phi$ and $\phi$ denote the Bardeen potential and an auxiliary
random-Gaussian field.

Primordial non-Gaussianity has primarily been constrained from the
bispectrum in cosmic microwave background (CMB) temperature fluctuations. 
Recently, Planck~\cite{Ade:2013ydc} obtained a tight constraint of
$f_{\rm NL}=2.7\pm 5.8$ at $1\sigma$ statistical significance. 
A complementary way to access non-Gaussianity is to measure its impact
on large scale structure. 
Luminous sources such as galaxies must be most obvious tracers 
of the underlying dark matter distributions with a bias.
Primordial non-Gaussianity induces the
scale-dependent bias~\cite{Dalal:2007cu,Desjacques:2008vf} such that
the effect dominates at very large scales. Hence, based on a
reasonable assumption that the galaxy bias is linear and deterministic
on large scales, it has been shown that the galaxy survey can
effectively constrain $f_{\rm NL}$ to the level comparable to CMB
temperature anisotropies~\cite{Ho:2013lda,Giannantonio:2013uqa}.
While clustering analyses at large scales are limited due to 
cosmic variance, Seljak~\cite{Seljak:2008xr} proposed a novel method
to reduce the cosmic variance using multiple tracers with different
biases, the so-called multitracer technique. This method allows us to
measure the scale-dependent bias accurately even at large scales,
leading to strong constraints on $f_{\rm NL}$. 

Future wide and deep surveys with 
Euclid\footnote{See http://www.euclid-ec.org} in optical and infrared bands
and Square Kilometre Array (SKA)~\footnote{See http://www.skatelescope.org}
in radio wavelengths will provide an unprecedented number of galaxies
to measure the power spectra. The radio continuum survey conducted
with SKA covers $30,000\,{\rm deg}^2$ out to high redshifts, though
the redshift information is not available. The authors in
\cite{Ferramacho:2014pua} found that even without the redshift
information the multitracer technique improves constraints as $\sigma
(f_{\rm NL})=\mcO (1)$, while weaker constraints of $\sigma (f_{\rm
  NL})=\mcO (10)$ without the multitracer technique. 
While the number of galaxies and covered area are smaller for the
Euclid photometric survey ($15,000\,{\rm deg}^2$), it provides
redshift information via photometric redshifts. Redshift information
is expected to be highly advantageous for constraining $f_{\rm NL}$
because the bias evolves strongly with redshift. As we show below, each of
these two surveys provides constraints of $\sigma (f_{\rm NL})=\mcO
(1)$\, and constraints improve to $\sigma (f_{\rm NL})=\mcO (0.1)$\,
with their combination. To calculate expected constraints, in this
paper, we employ the Fisher matrix formalism including the redshift
binning as well as the mass binning, taking the overlap of the two
survey regions into account. 

\section{Primordial non-Gaussianity in the large scale structure clustering} 

First, we consider the non-Gaussian correction of the halo bias given
by \cite{Desjacques:2008vf}
\al{
	\Delta b
		=&\frac{2f_{\rm NL}\delta_{\rm c}}{\mcM D_+}
				\left( b_{\rm L}-1\right)
				-\frac{1}{\delta_{\rm c}}\frac{\dd}{\dd\ln\nu}\left(\frac{\dd n/\dd M}{\dd n_{\rm G}/\dd M}\right)
	\,,
}
where $\nu =\delta_c /\sigma$\,, $\delta_{\rm c}\approx 1.68$ is
the critical linear density for spherical collapse and
$\sigma (M,z)=\sigma_R (z)$ is the variance of the linear density field
smoothed on the scale $R(M)=(3M/4\pi\rho_{\rm b,0})^{1/3}$
with $\rho_{\rm b,0}$ being the background density today.
$D_+(z)$ is the growth factor, 
$\mcM (k)=2k^2T(k)/3\Omega_{\rmm ,0}H_0^2$\,, where $T(k)$ is
the matter transfer function normalized
to unity at large scales~\cite{Eisenstein:1997ik}. 
We employ a fit to simulation for the Gaussian mass function
$\dd n_{\rm G}/\dd M$ and the linear bias factor
$b_{\rm L}$ given in \cite{Sheth:1999mn}.
We adopt a non-Gaussian correction of the mass function developed
in \cite{LoVerde:2007ri}, where we need the skewness of the density field
that is proportional to
$f_{\rm NL}$~\cite{Desjacques:2008vf,Oguri:2009ui,Chongchitnan:2010xz}.
In this paper, for $\sigma S_3$, we adopt a fitting formula from
\cite{Oguri:2009ui}.

Constraints on $f_{\rm NL}$ come from the redshift and mass dependences
of the bias. Thus, in order to take advantage of the multitracer
technique, we need a rough estimate of the halo mass of each galaxy. 
In the Euclid survey, assuming an accurate photometric redshift
estimate of each galaxy, we can use various galaxy properties such as
luminosity, color, and stellar mass to infer the halo mass.
See e.g. \cite{Vale:2004yt,2012ApJ...744..159L} for details.
On the other hand, it is more challenging to estimate the halo mass of
galaxies from radio surveys. In this paper, following
\cite{Ferramacho:2014pua}, we assume that halo mass can be estimated
from the galaxy type. 

Estimates of the halo mass for individual galaxies involve large
uncertainties. We take account of the uncertainties in halo mass
estimation following \cite{Lima:2004wn}. Given the estimated mass 
$M_{\rm est}$, the probability that the true mass is $M$ is assumed
to be given by log-normal distribution with the variance
$\sigma_{\ln M}^2$ and the bias $\ln M_{\rm bias}$\,,
\al{
	x(M_{\rm est};M)
		=\frac{\ln M_{\rm est}-\ln M-\ln M_{\rm bias}}{\sqrt{2}\sigma_{{\rm ln}M}}
	\,.
}
Furthermore, it is expected that these parameters depend on both halo
mass and redshift. We assume the following functional 
form~\cite{Oguri:2010vi,Cunha:2010zz}:
\al{
	&\ln M_{\rm bias}(M,z)
		=\ln M_{\rm b,0}		
	\notag\\
	&\quad\quad
			+\sum_{i=1}^3q_{{\rm b},i}
				\biggl[\ln\left(\frac{M}{M_{\rm piv}}\right)\biggr]^i
			+\sum_{i=1}^3s_{{\rm b},i}z^i
	\,,\label{eq:ln M_bias}\\
	&\sigma_{{\rm ln}M}(M,z)
		=\sigma_{{\rm ln}M,0}
	\notag\\
	&\quad\quad
			+\sum_{i=1}^3q_{\sigma_{{\rm ln}M},i}
				\biggl[\ln\left(\frac{M}{M_{\rm piv}}\right)\biggr]^i
			+\sum_{i=1}^3s_{\sigma_{{\rm ln}M},i}z^i
	\,,\label{eq:sigma_ln M}
}
with $M_{\rm piv}=10^{12}h^{-1}M_\odot$. Here we included a large
number of parameters that model the uncertainty of the halo mass
estimate, which are fully marginalized over when deriving constraints
on $f_{\rm NL}$. 

To apply the multitracer technique, we split galaxy samples
into $N_M$ mass-divided subsamples for each redshift bin. The average
density of galaxies in the $i$ th redshift bin $z_i<z<z_{i+1}$
and the $b$ th mass bin $M_{(b)}<M_{\rm est}<M_{(b+1)}$ is given by
\al{
	\bar N_{i(b)}
		=\int_0^\infty\dd z\frac{\dd^2 V}{\dd z\dd\Omega}
			\int_0^\infty\dd M\frac{\dd n}{\dd M}S_{i(b)}
	\,.\label{eq:bar N_ib}
}
Here $\dd^2 V/\dd z\dd\Omega =\chi^2 /H$ denotes the comoving volume
element per unit redshift per unit steradian, and we have introduced
$S_{i(b)}(M,z)$ to represent the selection function:
\al{
	&S_{i(b)}(M,z)=\Gamma_{(b)}\Theta (z-z_i)\Theta (z_{i+1}-z)
	\notag\\
	&\quad\times
					\frac{1}{2}
						\biggl[
							{\rm erfc}\Bigl( x(M_{(b)};M)\Bigr)
							-{\rm erfc}\Bigl( x(M_{(b+1)};M)\Bigr)
						\biggr]
	,
}
where we have introduced the gray-body factor $\Gamma_{(b)}$ to denote
the fraction of observed halos for each mass bin, since we may not be
able to observe all galaxies associated with the underlying dark
matter halos. 
With these variables, the Limber-approximated angular power spectrum
between $b$ and $b'$ th mass bins in the $i$ th redshift bin is expressed
by \cite{Oguri:2010vi}
\al{
	C_{i(bb')}(\ell )
		=&\int_0^\infty\dd zW_{i(b)}W_{i(b')}
		\frac{H}{\chi^2}P_\delta\left( \frac{\ell +1/2}{\chi},z\right)
	\,,\label{C_l def}
}
where $P_\delta (k,z)$ is the underlying dark matter power spectrum
and $W_{i(b)}$ is the weight function defined as
\al{
	W_{i(b)}
		=&\frac{1}{\bar N_{i(b)}}\frac{\dd^2 V}{\dd z\dd\Omega}
			\int_0^\infty\dd M\frac{\dd n}{\dd M}
			S_{i(b)}b_h\left( M,z,\frac{\ell +1/2}{\chi}\right)
	\,.\label{eq:W def}
}

\section{Fisher matrix formalism} 

We adopt the Fisher analysis to estimate expected errors of model parameters
for a given survey. The Fisher matrix is defined by
\al{
	F_{\alpha\beta}
		=\sum_{\ell =\ell_{\rm min}}^{\ell_{\rm max}}
			\sum_{I,J}
			\frac{\pd C_I(\ell )}{\pd\theta^\alpha}
			\Bigl[{\rm Cov}({\bm C}(\ell ),{\bm C}(\ell ))\Bigr]^{-1}_{IJ}
			\frac{\pd C_J(\ell )}{\pd\theta^\beta}
	\,,\label{eq:Fisher}
}
where the indices $I$ and $J$ run over the redshift and mass bin,
$(i,b,b')$\,, and $\theta^\alpha$ are model parameters.
Here, we consider $29$ parameters in the Fisher matrix analysis:
the primordial non-Gaussianity parameter $f_{\rm NL}$, 14 parameters
for systematic errors in the halo mass estimate for each of Euclid
and SKA [see Eqs.~\eqref{eq:ln M_bias} and \eqref{eq:sigma_ln M}].
We choose $\sigma_{\ln M,0}=0.3$ and zero for the other parameters
as fiducial values. On the other hand, we fix standard cosmological
parameters to those of the standard $\Lambda$CDM model: 
$\Omega_{\rm m,0}=0.266$\,, 
$\Omega_{\rm b,0}=0.04479$\,, 
$\Omega_\Lambda =0.734$\,,
$w=-1$\,,
$h=0.710$\,, 
$n_{\rm s}=0.963$\,,
$k_0=0.05{\rm Mpc}^{-1}$
and
$\sigma_8 =0.801$\,.
The marginalized error on each parameter is given by
$\sigma (\alpha )=\sqrt{({\bm F}^{-1})_{\alpha\alpha}}$\,.

Now we derive the covariance matrix generalized to multiple tracers
which are observed in different sky areas with some overlap.
We introduce the observed density contrast as
\al{
	\delta_w^{i(b)}({\bm\theta})
		= w_{(b)}({\bm\theta}) \delta^{i(b)}({\bm\theta})\,,
}
where $w_{(b)}({\bm\theta})$ is the survey window function on the sky
for $b$ th tracer; $w_{(b)}=1$ if the direction ${\bm\theta}$ on the sky
is in the survey region, otherwise $w_{(b)}=0$\,. With the two-dimensional
Fourier components of $\delta_w^{i(b)}({\bm\theta})$,
\al{
	\tilde{\delta}_w^{i(b)}({\bm\ell})
		= \int \dd^2{\bm\ell}'(2\pi )^{-2}
			\tilde{w}_{(b)}({\bm\ell}-{\bm\ell}')
			\tilde{\delta}^{i(b)}({\bm\ell}')
	\,,
}
where $\tilde{w}_{(b)}$ and $\tilde{\delta}^{i(b)}$ are Fourier transform of
$w_{(b)}$ and $\delta^{i(b)}$, respectively, we can define an estimator
of the angular power spectrum as \cite{Takada:2013wfa}
\al{
	\hat C_{i(bb')}(\ell )
		=\frac{1}{\Omega_w^{(bb')}}\int_{|{\bm\ell}'|\in\ell}\frac{\dd^2{\bm\ell}'}{\Omega_\ell}
			\tilde\delta_w^{i(b)}({\bm\ell'})\tilde\delta_w^{i(b')}(-{\bm\ell}')
	\,,
}
where we have considered the integral over a shell in the Fourier space of
width $\Delta\ell$ and volume
\al{
	\Omega_\ell
		= \int_{|{\bm\ell}|'\in\ell} \dd^2{\bm\ell}'\approx 2\pi\ell\Delta\ell
\,.
}
Here the effective survey area was defined as
\al{
	\Omega_w^{(bb')}=\int\dd^2{\bm\theta}w_{(b)}w_{(b')}
	\,,
} 
which is
the survey area of the $b$ th tracer for $b=b'$ and the overlapping area
of the $b$ and $b'$ th tracers for $b \neq b'$.
We have determined the functional form of the estimator so that it is
unbiased in a sense that the ensemble average gives the true power spectrum,
namely, $\big\langle\hat C_{i(bb')}(\ell )\big\rangle = C_{i(bb')}(\ell )$\,.
Assuming the Gaussian error covariance, we obtain the covariance matrix
for multiple tracers as
\al{
	&{\rm Cov}\Bigl[C_{i(bb')}(\ell ) ,C_{j(\tilde b\tilde b')}(\ell' )\Bigr]
		=\frac{\delta_{ij}^{\rm K}\delta_{\ell\ell'}^{\rm K}}{(2\ell +1)\Delta\ell}
			\frac{4\pi\Omega_w^{(bb'\tilde b\tilde b')}}{\Omega_w^{(bb')}\Omega_w^{(\tilde b\tilde b')}}
	\notag\\
	&\quad\times
			\biggl[
				C_{i(b\tilde b)}(\ell )C_{i(b'\tilde b')}(\ell )
				+C_{i(b\tilde b')}(\ell )C_{i(b'\tilde b)}(\ell )
			\biggr]
	\,,\label{eq:covariance matrix}
}
with
\al{
	\Omega_w^{(bb'\tilde b\tilde b')}
		= \int \dd^2{\bm\theta} w_{(b)} w_{(b')} w_{(\tilde b)} w_{(\tilde b')}
	\,.
}
Since the observed spectrum includes the shot noise contamination,
we replace $C_{i(bb')}(\ell )$ with
$C_{i(bb')}(\ell ) + \bar{N}_{i(b)}^{-1}\delta_{bb'}^{\rm K}$\,.

\begin{figure}[tbp]
\bc
\includegraphics[width=70mm]{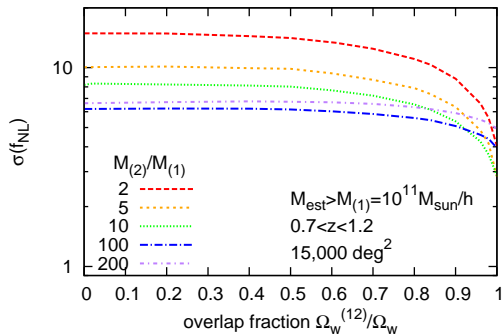}
\caption{
The marginalized error on $f_{\rm NL}$
as the function of the overlap fraction, for the single redshift bin
of $0.7<z<1.2$.  Different lines show results with different 
mass ratio $M_{(2)}/M_{(1)}$.
}
\label{fig:sigma_fNL_fSkyOL}
\ec
\end{figure} 

\section{Results} 

As we stated above, we consider the Euclid photometric survey and
the SKA continuum survey. For Euclid, a redshift range $0.2<z<z_{\rm max}$ is
considered and galaxy samples are split into several redshift bins with
the same interval ($\Delta z=0.5$). We neglect the photometric
redshift errors as they are expected to be much smaller than 
$\Delta z$. To include the effect of flux cut for each redshift range, 
we adopt the following minimum observed mass for each bin,
$M_{\rm est}>0.7\,,1\,,2\,,5\,,10\,,20\,,50\,,100\,,\cdots$
in the unit of $10^{11}h^{-1}M_\odot$\, and set
$\Gamma_{(b)}^{\rm Euclid}=1$\,. Galaxy samples are further split
according the estimated halo mass. We consider five mass-bins and
take separating masses such that the five mass bins of the same
redshift bin have the same number of samples. Here it should be noted
that the separating masses depend on the redshift.
We will discuss other possibilities of the mass binning later. 
Summation of the power spectrum is taken for an $\ell$ range of
$3 \leq \ell \leq 400$\,.

As for the SKA continuum survey, we have only one redshift bin as no
redshift information is available. Thus we simply drop the redshift
dependent terms in Eqs.~\eqref{eq:ln M_bias} and \eqref{eq:sigma_ln M}.
Following \cite{Ferramacho:2014pua}, we consider five types of galaxies
as five tracers with the typical masses \cite{Wilman:2008ew},
$M_{\rm SFG}=10^{11}h^{-1}M_\odot$ for star forming galaxies,
$M_{\rm RQQ}=3\times 10^{12}h^{-1}M_\odot$ for radio quiet quasars,
$M_{\rm FRI}=10^{13}h^{-1}M_\odot$ for FRI,
$M_{\rm SB}=5\times 10^{13}h^{-1}M_\odot$ for starburst galaxies
and $M_{\rm FRII}=10^{14}h^{-1}M_\odot$ for FRII.
Accordingly, we consider five mass bins,
$M_{(i)} < M < M_{(i+1)} ~ (i=1,\cdots, 4)$ and $M > M_{(5)}$, with
$M_{(1)}=0.9\times 10^{11}h^{-1}M_\odot$\,, 
$M_{(2)}=\sqrt{M_{\rm SFG}M_{\rm RQQ}}$\,, 
$M_{(3)}=\sqrt{M_{\rm RQQ}M_{\rm FRI}}$\,,
$M_{(4)}=\sqrt{M_{\rm FRI}M_{\rm SB}}$\,, 
$M_{(5)}=\sqrt{M_{\rm SB}M_{\rm FRII}}$\,.
For the flux cut, we adopt the gray body factor as
$\Gamma_{(b)}^{\rm SKA1}=\{0.013,0.03,0.1,1,1\}$
and $\Gamma_{(b)}^{\rm SKA2}=\{0.5,1,1,1,1\}$, which are chosen to
match the expected number density distribution of galaxies found in
these surveys (see e.g., \cite{Ferramacho:2014pua}). As for
$\ell$ range, we consider $2 \leq \ell \leq 400$\,. 

In computing the Fisher matrix for the combination of Euclid and SKA surveys,
we adopt $9,000\,{\rm deg}^2$ as the area of the overlap region 
and we neglect the contributions from the derivative of
the cross correlations between Euclid and SKA for simplicity.
We focus on constraints on $f_{\rm NL}$ and marginalize over the other
parameters. 

Before showing expected constraints from Euclid and SKA surveys,
let us check the dependence of the efficiency of the multitracer
technique on the overlapping survey area and different mass binning,
considering a simple case of two tracers observed by a Euclid-like
survey. In Fig.~\ref{fig:sigma_fNL_fSkyOL}, we plot the marginalized
error on $f_{\rm NL}$ as a function of the overlap fraction
$\Omega_w^{(12)}/\Omega_w$ for a single redshift bin
$0.7<z<1.2$\,. Different curves represent different mass binning 
varying the mass ratio $M_{(2)}/M_{(1)}$\,. Here we assume that
the sky coverages for both tracers are the same,
$\Omega_w^{(11)}=\Omega_w^{(22)}\equiv\Omega_w$.
We find that the nonvanishing overlap region leads to improved constraints
on $f_{\rm NL}$\,, which becomes smallest in the case of the maximal overlap.
One can also see that in the case of the maximal overlap there is
a critical value of the mass ratio $M_{(2)}/M_{(1)}$ which results in
the tightest constraint. This behavior can be understood as follows:
once we fix the mass ratio, the number density for each mass bin,
$\bar N_{i(b)}$, is determined through Eq.~\eqref{eq:bar N_ib}\,. 
Changing the value of the mass ratio leads to the larger shot noise
for one of the mass-bins and smaller shot noise for the other.
We find that the tightest constraint is obtained when the shot noise
for the two mass bins becomes comparative. This is the reason for our
choice of separating masses by the same number density, as explained
above. 

\begin{figure}[tbp]
\bc
\includegraphics[width=70mm]{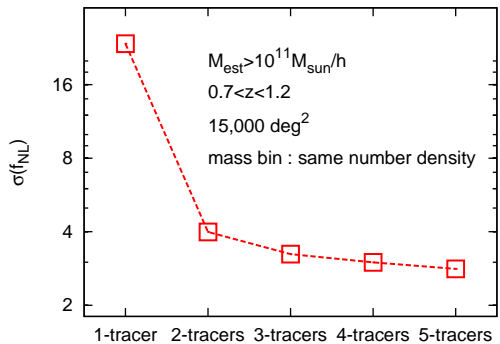}
\caption{
The marginalized error on $f_{\rm NL}$
as the function of the number of the tracers
in the single redshift bin $0.7<z<1.2$.  The mass bins are divided
such that they have the equal shot noises. 
}
\label{fig:sigma_fNL_nTracer}
\ec
\end{figure} 

\begin{figure}[tbp]
\bc
\includegraphics[width=70mm]{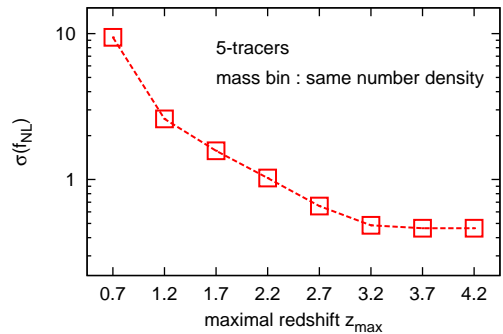}
\caption{
The marginalized constraint on $f_{\rm NL}$ as a function of
the maximum redshift, assuming the redshift range $0.2 < z < z_{\rm max}$
with width $\Delta z=0.5$. Here we take five tracers (mass bins) for each
redshift bin.
}
\label{fig:sigma_fNL_zbin}
\ec
\end{figure}

Next, we focus on the Euclid survey. Figure \ref{fig:sigma_fNL_nTracer} shows
the marginalized constraints on $f_{\rm NL}$ as a function of the number
of tracers for a single redshift bin $0.7<z<1.2$ with the maximal overlap
among tracers. We find that the constraining power increases with $N_M$.
Even 2 tracers drastically improve the constraint, simply because the
multitracer technique does not take effect for the one tracer case.
Furthermore, combining multiple redshift bins improves substantially
the constraint, as is shown in Fig.~\ref{fig:sigma_fNL_zbin}. We
find that galaxy samples as far as $z = 3.2$ (sixth bin) contribute
significantly to the constraint. When five mass bins and eight
redshift bins are taken into account, the Euclid photometric survey
can reach $\sigma (f_{\rm NL})=0.46$\,.
Although the use of galaxies out to $z=4.2$ is probably too optimistic,
even in a more realistic situation where we use redshift bins up to
$z=2.7$ (five bins) the improvement is still significant, $\sigma (f_{\rm NL})=0.66$\,.
In the reminder of the paper we conservatively adopt $z_{\rm max} = 2.7$ as
the maximal redshift for Euclid.

\begin{figure}[tbp]
\bc
\includegraphics[width=70mm]{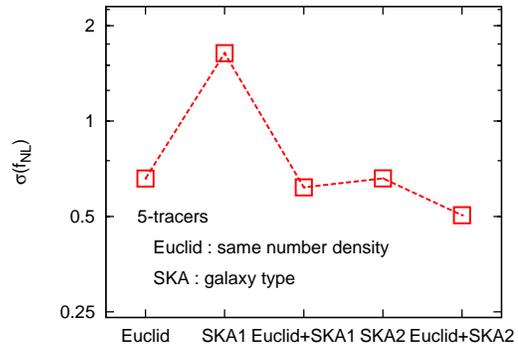}
\caption{
The expected marginalized constraints on $f_{\rm NL}$
for each survey and combinations.
}
\label{fig:sigma_fNL_survey}
\ec
\end{figure} 

Finally, Fig.~\ref{fig:sigma_fNL_survey} shows 
the expected marginalized constraints on $f_{\rm NL}$ for each survey
and their combinations. The constraints on $f_{\rm NL}$ from SKA1 and
SKA2 are $\sigma (f_{\rm NL})=1.64$\,, $0.66$\,, respectively, which are
consistent with Ref.~\cite{Ferramacho:2014pua}.
The results of SKA2 and SKA1 are comparable to or relatively weaker 
than that from Euclid, presumably because
the redshift information obtained from the photometric survey is more
advantageous than the larger sky coverage and the larger number 
of galaxy samples from SKA survey. Combining Euclid and SKA,
the constraint can improve further to $\sigma (f_{\rm NL})=0.61$
(Euclid+SKA1)\,, $0.50$ (Euclid+SKA2), suggesting that the joint analysis
between Euclid and surveys are quite effective to constrain primordial
non-Gaussianity.
We again note that the improvement of the constraint is mainly
due to the availability of the multiple tracer, as seen in Fig.~\ref{fig:sigma_fNL_nTracer}.
Although the results presented would be sensitive to the assumptions we considered
in this paper, the constraining power due to the multitracer technique is expected to be generic 
and the behavior of the results would remain the same.

\section{Summary} 

To summarize, we have discussed the potential power of multitracer
technique for the combination of the Euclid photometric survey and the
SKA continuum survey. Splitting the galaxy samples into the subsamples
by the inferred halo mass and redshift, constraints on $f_{\rm NL}$
drastically improve. We have shown that constraints of $\sigma (f_{\rm
  NL}) = \mcO(1)$ can be obtained even with a single survey. Combining
Euclid and SKA, even stronger constraints of $\sigma (f_{\rm NL}) =
\mcO(0.1)$ can be obtained. 

In this paper, we have made several simplified assumptions.
In future galaxy surveys, the systematic uncertainties likely play a
more important role than statistical errors. Here we considered only
the uncertainty in the halo mass estimation. For instance, the
uncertainty in photometric redshifts and the effect of the stochastic
bias may become important. 
We should also address the identification of the optical and infrared counterparts
in the overlap region of SKA and Euclid surveys. While we
conservatively assumed no redshift information for the SKA survey,
checking the counterparts in Euclid or other surveys
would provide valuable information on redshifts of individual SKA
sources, which may allow the tomographic analysis in 
the SKA survey to lead further improvements of the constraints
(see \cite{Raccanelli:2014kga}). We hope to come back these issues in
the near future.


\acknowledgments
D.Y. is supported by Grant-in-Aid for JSPS
Fellows (No.~259800). This work was supported by Grant-in-Aid from the
Ministry of Education, Culture, Sports, Science and Technology (MEXT)
of Japan, No.~24340048, (K.T.), No.~26610048 (K.T.), and No.~26800093 (M.O.).


\end{document}